# Possible signatures for strange stars in stellar X–ray binaries

Bhaskar Datta[1,2], Arun V. Thampan[3], and Ignazio Bombaci[4]

[1] Indian Institute of Astrophysics, Bangalore 560 034, INDIA
(datta@iiap.ernet.in)

[2] Raman Research Institute, Bangalore 560 080, INDIA

[3] Inter-University Centre for Astronomy and Astrophysics (IUCAA), Pune 411 007, INDIA
(arun@iucaa.ernet.in)

[4] Dipartimento di Fisica Universitá di Pisa, and INFN Sezione di Pisa, via Buonarroti 2, 56127 Pisa, ITALY
(BOMBACI@pi.infn.it)

**Abstract.** Kilohertz quasi–periodic brightness oscillations (kHz QPOs) observed in certain X–ray burst sources may represent Keplerian frequencies in the inner regions of the accretion disk in such systems. If this assumption is strictly adhered to, we show here that if the central accretor in stellar X–ray burst sources is a strange star (made up of u, d and s quarks in beta equilibrium, referred to as strange matter) then the calculated QPO frequencies are reconcilable with the observed QPO frequencies (corresponding to the highest frequency of 1.22 kHz, observed so far from the source 4U 1636–53) only for particular values of the QCD-related parameters which describe the equation of state of strange matter. We demonstrate that QPO frequencies in the very high range (1.9–3.1) kHz can be understood in terms of a (non–magnetized) strange star X–ray binary (SSXB) rather than a neutron star X–ray binary (NSXB). Future discovery of such high frequency QPOs from X–ray burst sources will constitute a new astrophysical diagnostic for identifying solar mass range stable strange stars in our galaxy.

## 1. Introduction

Discovery of kHz QPOs in the flux from certain X–ray burst sources have prompted substantial amount of work in connection with accretion physics and structure properties of the central accretors in such systems. In particular, these oscillations have been used to

---

derive estimates of the mass of the neutron star in X–ray binaries (Kaaret, Ford & Chen 1997; Zhang, Strohmayer & Swank 1997; Kluźniak 1998). All these estimates, based on the beat frequency model, tacitly assume that the highest QPO frequency of 1.22 kHz observed so far (in the source 4U 1636–53; Zhang et al. 1997) can be identified with the Keplerian orbital frequency corresponding to the marginally stable orbit associated with the neutron star. Beat frequency models require that the difference in frequencies between the twin QPO peaks be the spin frequency of the neutron star and that this remain constant. However, further observations have shown that there exist microsecond lags in the QPO difference frequencies in many sources implying that an exact beat frequency mechanism may not be at work. Recently, Osherovich & Titarchuk (1999a), Titarchuk & Osherovich (1999), Osherovich & Titarchuk (1999b) have developed alternative models unifying the mechanism for production of low frequency QPOs and that for high frequency QPOs. This model requires the lower frequency QPO to be due to Keplerian circulation of matter in the disk and the higher frequency one to be hybrid between the lower frequency and the rotational frequency of the stellar magnetosphere. Li et al. (1999b) have suggested that if such a model is taken recourse of, then the compact star in the source 4U 1728 – 34 may possibly be a strange star.

The possible existence of a new sequence of degenerate compact stellar objects, made up of light mass u, d and s quarks, has been suggested (Witten 1984; Haensel, Zdunik & Schaeffer 1986; Alcock, Farhi & Olinto 1986) for quite sometime now, based on ideas from particle physics which indicate that a more fundamental description of hadronic degrees of freedom at high matter densities must be in terms of their quark constituents. For energetic reasons, a two–component (u,d) quark matter is believed to convert to a three–component (u,d,s) quark matter in beta equilibrium. As suggested by Witten (1984), the latter form of matter could be the absolute ground state of strongly interacting matter rather than $^{56}Fe$. Because of the important role played by the confinement forces in quantum chromodynamics (QCD) to describe the quark interactions, the mass–radius relationship for stable strange stars differ in an essential manner from that of neutron stars (Haensel, Zdunik & Schaeffer 1986; Alcock, Farhi & Olinto 1986). Recent work (Cheng et al. 1998; Li et al. 1999a, Li et al. 1999b) seem to suggest that a consistent explanation of the observed features of the hard X–ray burster GRO J 1744 – 28, the transient X–ray burst source SAX J 1808.4 – 3658 and the source 4U 1728 – 34 is possible only in terms of an accreting strange star binary system. A new class of low–mass X–ray binaries, with strange star as the central compact object (SSXBs), is thus an interesting astrophysical possibility that merits study. Some consequences of the SSXB hypothesis for the properties of bulk strange matter have been discussed recently by Bulik, Gondek-Rosińska and Kluźniak (1999) (see also Schaab & Weigel 1999).

The compact nature of the sources make general relativity important in describing these systems. Furthermore, their existence in binary systems imply that these may possess rapid rotation rates (Bhattacharya & van den Heuvel 1991 and references therein). These two properties make the incorporation of general relativistic effects of rotation imperative for satisfactory treatment of the problem. General relativity predicts the existence of marginally stable orbits around compact stars. For material particles within the radius of such orbits, no Keplerian orbit is possible and the particles will undergo free fall under gravity. This radius ($r_{ms}$) can be calculated for equilibrium sequences of rapidly rotating strange stars in a general relativistic space–time in the same way as for neutron stars (Datta, Thampan & Bombaci 1998).

In this letter, we calculate the Keplerian frequency of matter revolving around rapidly rotating strange stars. The present results, together with those obtained assuming a neutron star as the central accretor (Thampan et al. 1999), demonstrate that QPO frequencies in the range (1.9-3.1) kHz can be interpreted in terms of a non-magnetized SSXB rather than a NSXB. Future discovery of such high frequency QPOs from X–ray burst sources will constitute a new astrophysical diagnostic for SSXBs. In section (2) we very briefly discuss the formalism used to construct rapidly rotating strange star sequences and further computing the Kepler frequencies around such objects. Section(3) provides a brief outline of the equation of state (EOS) models used by us. In section (4) we discuss the results and conclusions.

## 2. Calculations

We use the methodology described in detail in Datta, Thampan & Bombaci (1998) to calculate the structure of rapidly rotating strange stars. For completeness, we briefly describe the method here. For a general axisymmetric and stationary space–time, assuming a perfect fluid configuration, the Einsten field equations reduce to ordinary integrals (using Green's function approach). These integrals may be self consistently (numerically and iteratively) solved to yield the value of metric coefficients in all space. Using these metric coefficients, one may then compute the structure parameters, moment of inertia and angular momentum corresponding to initially assumed central density and polar to equatorial radius ratio. The values of the structure parameters and the metric coefficients, so computed, may then be used (as described in Thampan & Datta 1998) to calculate parameters connected with stable circular orbits (like the innermost stable orbit and the Keplerian angular velocities) around the configuration in question.

## 3. Strange star equations of state

For purpose of this letter, we have calculated the relevant quantities (of interest here), corresponding to three different equation of state (EOS) models for strange stars. Two of these equations of state are based on the MIT bag model (Chodos et al. 1974) with the following values for the bag pressure ($B$), the strange quark mass ($m_s$) and the QCD structure constant ($\alpha_c$): (i) $B = 90$ MeV fm$^{-3}$, $m_s = 0$ MeV and $\alpha_c = 0$; (ii) $B = 56$ MeV fm$^{-3}$, $m_s = 150$ MeV, with the short range quark–quark interaction incorporated perturbatively to second order in $\alpha_c$ according to Freedman & McLerran (1978) and Goyal & Anand (1990). Next we considered a phenomenological model by Dey et al. (1998) (model (iii)) that has the basic features of QCD (namely, quark confinement and asymptotic freedom), but employs a potential description for the interaction. These models for the EOS are quite divergent in their approach, so that the conclusions presented here using these will be of sufficient generality.

## 4. Results and Conclusions

For the EOS models described in the previous section, we calculate the Keplerian frequencies corresponding to the innermost 'allowed' orbits (as given by general relativity) for rotating strange stars, and obtain their relationship with QPO frequencies in the kHz range, assuming the SSXB scenario. The inner edge of the accretion disk may not always be coincident with $r_{ms}$, but there can be instabilities in the disk that can relocate it outside of $r_{ms}$. If the radius ($R$) of the strange star is larger than $r_{ms}$, the innermost possible orbit will be at the surface of the strange star. It must be mentioned here that rotation of the central accretor is an important consideration because the accretion driven angular momentum transfer over dynamical timescales can be quite large (Bhattacharya & van den Heuvel 1991). Because the values of $r_{ms}$ and the mass of the spinning strange star will depend on two independent parameters, namely, the central density ($\rho_c$) of the star and its spin frequency ($\nu_s$), a range of values of ($\rho_c,\nu_s$) will exist that will allow solutions for a Keplerian frequency corresponding to any specified value of the QPO frequency.

The variation of the Keplerian frequency ($\nu_K$) of the innermost 'allowed' orbit with respect to the gravitational mass (M) of the spinning strange star is shown in Fig. 1. For purpose of illustration, we have chosen three values of $\nu_s$ : 0 (the static limit), 200 Hz and 580 Hz (the last rotation rate inferred from the X–ray source 4U 1636–53 as given by Zhang et al. 1997, using beat frequency model). It can be noted from Fig. 1 that all the curves have a cusp. For any curve, the nearly flat part (to the left of the cusp) corresponds to the case $R \geq r_{ms}$, and the descending part (to the right of the cusp) corresponds to the case $R \leq r_{ms}$. These are the only possibilities for the location of $r_{ms}$ with respect to the stellar surface. The highest kHz QPO frequency observed so far is 1.22 kHz, exhibited by the source 4U 1636–53. Fig. 1 shows that only the maximum mass end of the curve for non–rotating configuration described by EOS model (ii) attains the value $\nu_K = 1.22$ kHz. A simple analysis, relating the minimum value of $\nu_K$ to the bag constant (see Fig. 1 for EOS (i)) in the case of non-rotating strange stars within the MIT bag model EOS for massless non-interacting quarks gives $\nu_K(r_{ms}, M_{max}) = 1.081(B/56)^{1/2}$ kHz, where B is in MeV fm$^{-3}$. The lowest possible value for $B$, which is compatible with the Witten's hypothesis (Witten 1984), is 56 Mev fm$^{-3}$. Finite values of $m_s$, $\alpha_c$, and $\nu_s$ increase the value of $\nu_K(r_{ms}, M_{max})$ with respect to the previous case. This implies that, if one adheres to the restrictive assumption that $\nu_{QPO} = 1.22$ kHz in the X–ray source 4U 1636–53 is generated at the marginally stable orbit of the central compact star (with $r_{ms} > R$), then the latter being a strange star is an admissible solution only for low values of the bag constant and for very slowly rotating configurations of the star.

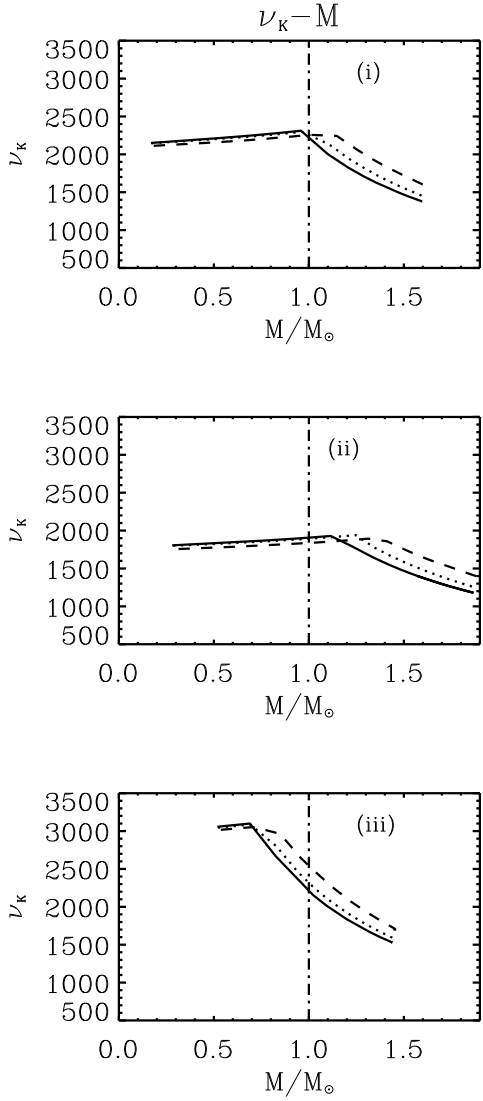

**Fig. 1.** The Kepler frequency $\nu_K$ corresponding to the innermost 'allowed' orbit as a function of gravitational mass $M$ of the neutron star. The three curves: solid, dotted and dashed are, respectively, for three values of neutron star spin frequency $\nu_s$, namely, 0, 200 and 580 Hz. The vertical dot–dashed line represents a 1 $M_\odot$ configuration. Each panel correspond to one of the EOS models described in the text.

Next we investigate the possibility that the kHz QPO frequency is generated at locations outside the marginally stable orbit. Since $\nu_K(r)$ is a decreasing function of r, $\nu_K = 1.22$ kHz in SSXBs will occur at $r > r_{ms}$, that is, somewhere in the accretion disk and not at the disk inner edge. In Fig. 2 we show the plot of the Keplerian frequency profiles $\nu_K(r)$ of test particles around a (rotating) strange star of one solar mass (for the same values of the rotation rates as before). This figure shows that the radial location in the disk, where a solution : $\nu_K = 1.22$ kHz occurs in a SSXB, is about $4.5 r_g$, where $r_g = 2GM/c^2$ is the Schwarzschild radius of the strange star. A similar analysis for $M = 1.4\ M_\odot$ yields $r_{1.22}$ (radius at which $\nu_{QPO} = 1.22$ kHz is produced) in the range (3.53, 3.55) $r_g$, the higher value being that for the non–rotating configuration and the lower for $\nu_s = 580$ Hz.

It is interesting to ask what range of $\nu_K$ obtains for a specified value of the strange star mass. From Fig. 1, it can be seen that the values of $\nu_K$ for SSXB, for a one solar mass strange star, lie in the range (2.2–2.3) kHz for EOS model (i), (1.8–1.9) kHz for EOS model (ii) and (2–2.6) kHz for EOS model (iii). The first two ranges of kHz QPOs occur at $r = R$, while the third at $r = r_{ms}$. For $M = 1.4\ M_\odot$, these ranges are: (1.57–1.84), (1.57-1.87) and (1.57–1.79), respectively for EOS models (i), (ii) and (iii). The similarity in these ranges is due to $r_{ms} > R$ for all these configurations. It also follows from Fig. 1 that the EOS model (iii) gives the maximum value of $\nu_K$, namely, 3 kHz.

The most interesting result ensues if a comparison is made of Fig. 1 with its counterpart for the case of a NSXB. A detailed calculation of the latter was reported recently by Thampan, Bhattacharya & Datta (1999), using realistic EOS models. This calculation showed that the maximum theoretically expected value of $\nu_{QPO}$ for NSXBs is 1.84 kHz. Therefore, values of $\nu_{QPO}$ in excess of 1.84 kHz, if observed, cannot be understood in terms of a NSXB. The SSXB scenario is a more likely one for these events (assuming that generation of X–ray bursts is possible on strange star surfaces); this will constitute a new astrophysical diagnostic for the existence of strange stars in our galaxy.

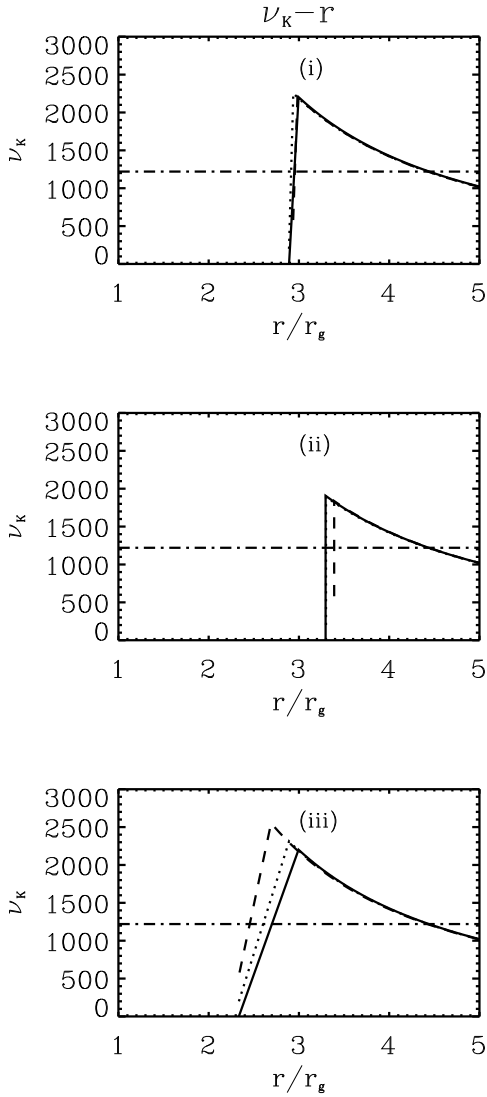

**Fig. 2.** Radial variation of $\nu_K$ for a 1 $M_\odot$ configuration. On the x–axis is the radial distance ($r$) scaled with the Schwarzschild radius ($r_g = 2GM/c^2$). The various curves have the same meaning as in Fig. 1. Where the dotted/dashed curves are not visible, they merge with the solid curve for the non–rotating configuration. The horizontal dot–dashed curve corresponds to $\nu_K = 1220 Hz$, the highest QPO frequency observed to date from the X–ray source 4U 1636−53. The $\nu_K = 1220$ Hz line intersects the curves (in all cases) at $r = 4.5 r_g$.